\documentclass[graphicx,aps,pre,epsfig,showpacs]{revtex4}
\usepackage{latexsym}
\usepackage{graphicx}
\usepackage{amsmath}

\def\ud{\mathrm{d}}
\def\B{\beta}
\def\hatu{\hat{u}(\underline{J},\underline{h})}
\def\hata{\hat{a}(\underline{J},\underline{h})}
\def\hatuu{\hat{u}(\underline{J},\underline{\sum^{K} u})}
\def\hatau{\hat{a}(\underline{J},\underline{\sum^{K} u})}

\begin{document}
\title{Zero temperature solutions of the Edwards-Anderson model in
  random Husimi lattices}

\author{Alejandro Lage-Castellanos }
\email{ale.lage@gmail.com}
\author{Roberto Mulet}
 \email{mulet@fisica.uh.cu}
\affiliation{Henri-Poincar\'e Group of Complex Systems, 
Physics Faculty, University of Havana, La Habana, CP 10400, Cuba}
\affiliation{
Department of Theoretical Physics, Physics Faculty, University of 
Havana, La Habana, CP 10400, Cuba}

\date{\today}

\begin{abstract}
We solve the Edwards-Anderson model (EA) in different Husimi
lattices using the cavity method at replica symmetric (RS) and 1-step of replica symmetry breaking (1RSB) levels. We show that, at $T=0$, the structure of the solution space depends on the parity of the loop sizes. Husimi lattices with odd loop sizes may have a trivial paramagnetic solution thermodynamically relevant for highly frustrated systems while, in Husimi lattices with even loop sizes, this solution is
 absent.  The range of stability under 1RSB perturbations of this and other RS solutions is computed analytically (when possible) or numerically. We also study the transition from 1RSB solutions to paramagnetic and ferromagnetic RS solutions. Finally we compare the solutions of the EA model in Husimi lattices with that on the (short loops free) Bethe lattices, showing that already for loop sizes of order 8 both models behave similarly.
\end{abstract}

\maketitle

\section{Introduction}
\label{sec:int}

Spin glasses are among the most complex problems in Statistical Mechanics. During the 80's a lot of effort was devoted to the subject, see for example \cite{MPV} for a comprehensive collection of relevant works during this decade. It was soon realized that the difficulties in finding an analytical solution to these problems depend strongly on the topology of the interactions between the variables \cite{MPV,VB}. For example, in fully connected systems, in which each variable interacts with all the others, a compact solution may be found using the Parisi ansatz\cite{MPV, a1, b1}. On the other hand, the situation for finite connectivity (FC) systems is far more complicated. The main difficulty is the appearance, after a standard replica calculation of an infinite number of overlaps\cite{DG,VB}. Therefore, for many years only Replica Symmetric or variational solutions were known for these models. 

The importance of finite connectivity systems is twofold: first, one may hope to get a better understanding of finite dimensional systems, since (FC) models include the notion of neighborhood, a concept that is absent in fully-connected systems. Second, there is a clear connection between finite connectivity systems and many constraint satisfaction problems. For example, the K-sat \cite{KS}, the coloring \cite{Col}, the traveling salesman \cite{TS} and the vertex cover \cite{VC1,VC2} problems turn out to have a finite connectivity structure.

A few years ago, Mezard and Parisi\cite{MP1,MP2} generalized a technique 
already known as the cavity method\cite{Cavold} to deal with systems with many
pure states. This generalization permitted, for the first time in FC
systems, the
formal introduction of replica symmetry breaking at different
levels. Although, it is worth reminding that even the 
one-Step Replica Symmetry Breaking solution (1RSB) involves as an order parameter a functional distribution.

Moreover, thanks to the aforementioned  strong connection between
finite connectivity spin systems and many constraint-satisfaction
problems\cite{review} this approach sheds some light on the
characteristics of the solution space of some of these problems
\cite{KS,Col,VC1,VC2}, (see also \cite{MRT,LK} for recent developments in this
field). In addition, this cavity method, inspired a novel message-passing algorithm to deal with single instances of several combinatorial problems \cite{KS2,Col2,VC3}. Unfortunately, the occurrence of short loops in finite connectivity
graphs introduces strong correlations among neighboring sites and in
this case, the hypothesis behind the cavity method may be violated. As
a consequence, message-passing algorithms usually fail when short
loops are present\cite{yedidia,MR,bolos1,bolos2}.
On the other hand, fully understanding the role of short loops in the energy landscape of finite dimensional spin glasses remains an elusive task\cite{MaPaRi,NS}.

To gain some insight about these problems, we think that it is convenient to look at the properties
of a spin glass model, where the hypothesis behind the cavity
approximation remain valid, but where the influence of short loop
structures may be analyzed in detail. Then, we choose to study the
ground state characteristics of the Edwards-Anderson model in a Husimi
graph. The  tree-like structure of the Husimi lattice allows 
the use of the machinery behind the cavity method, and at the same time, we may tune the loop sizes and the connectivity of the graph to discover their influence on the zero temperature energy landscape of the model.

The paper is organized as follows: In section
\ref{sec:mod} we introduce the model, the
self-consistent cavity equations that solve it and the definition of
the relevant physical quantities. Then, in section
\ref{sec:triang} we show closed analytical results for the triangular Husimi
lattice. In section \ref{sec:gene} appears the analysis of
more general lattices and 
finally in section \ref{sec:dis} we present the conclusions of the work.

\section{The model and the Cavity Solution}
\label{sec:mod}

\begin{figure}[htb]
        \begin{center}              
             \includegraphics[scale=0.55]{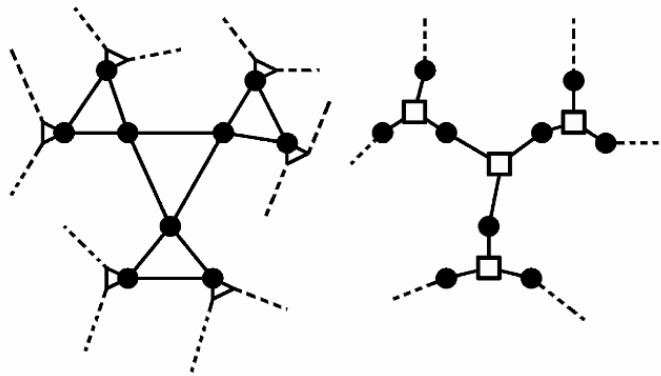} 
            \end{center} 
	\caption{Left: schematic representation of a representative part of a Husimi graph with
          $c=2$(triangle) and $K=1$. Right: factor graph
          representation of the same Husimi lattice. In the factor graph representation, $c+1$ is the degree of factor nodes (white squares), while $K+1$ is the degree of nodes (black circles).}
\label{fig:cactusFactorGraph}
 \end{figure}

A Husimi {\em tree} is formally ``a connected graph where no
bond (edge) lies in more than one cycle''. It can be visualized as a
tree made out of loops, as shown in figure \ref{fig:cactusFactorGraph}. A Husimi
tree is called {\em pure} if all the loops have the same length. It is also called
{\em regular} if all the vertexes belong to the same number of loops \cite{Ostilli}. On the
other hand, the
Husimi {\em lattice} (also Husimi {\em graph}) is a graph that looks locally like
a Husimi tree, but where large loops are present. It is a random
hyper-graph in the sense that by taking the short loops of the Husimi
tree as building units, we define the ensemble
of pure and regular Husimi graphs as the ensemble of all graphs in which the vertexes
belong to $K+1$ short loops of the same length $c+1$.

For example, the simplest (and trivial) Husimi tree is the one in which $c=1$ and coincides with
a Cayley tree. The simplest Husimi lattice, $c=1$ coincides with the
Bethe lattice defined in \cite{MP1}. The simplest, non-trivial, case
of a Husimi lattice has $c=2$, (a triangle) and $K=1$,  each vertex is
shared by two triangles (see figure \ref{fig:cactusFactorGraph}). The generalization to more complex structures
is straightforward, for example, either $c$ or $K$ or both may be taken as
random variables such that the local structure of the graph changes from
site to site. Here we consider the ensemble of pure and regular Husimi graphs, meaning
that all short loops have the same length, and that all vertexes belong to the same
number of (short) loops. 

In this work we study,  using the
cavity method at zero temperature \cite{MP2}, the ground state properties of the Edwards-Anderson model in an average Husimi graph. The Edwards-Anderson model is defined
by the Hamiltonian:

\begin{equation}
\mathcal{H}=-\sum_{<i,j>}J_{i,j}S_i S_j\label{eq:hamiltonianEA}
\end{equation}

\noindent where $<i,j>$ stands for the nearest neighbors in the graph,
$S_i=\pm 1$ are Ising variables located at the vertexes, and the bonds
represent random exchange couplings $J_{i,j}$.  The couplings are
taken from the distribution:

\begin{equation}
P(J)=\frac{1+\rho}{2}\delta(J-1)+\frac{1-\rho}{2}\delta(J+1)
\end{equation}

\noindent where  $\rho \in [-1,1]$ parameterizes the bias  between ferromagnetic
and anti-ferromagnetic interactions. If $\rho=1 (-1)$ the system is purely
ferromagnetic (anti-ferromagnetic), and if $\rho=0$ the system is unbiased.

As will be discussed in
more detail below, the cavity method may be easily written in
term of messages between Husimi {\em loops} and sites. The idea is to
use a factor graph that considers all the interactions around a loop as a
single function node (see figure \ref{fig:cactusFactorGraph}), then
inheriting the tree-like structure of the Husimi graph at 
the loop level. We will use the cavity method at two levels of
approximations: the replica symmetric approximation (RS) and the one
step replica symmetry breaking approximation (1RSB). It will turn out
that depending on $c$, $K$ and $\rho$, either the RS approximation or the 1RSB
are more appropriate to describe the ground state of the system. We
leave for future works the analysis of the stability of the 1RSB solution.

\subsection{The Self-Consistent Cavity Equations at $T=0$}

In order to write the cavity equations for the model, it is convenient to define
the problem using a factor graph representation. A factor
graph\cite{Ksch} is a bipartite graph in which one subset of the nodes represents
local interactions while the other subset represents variables. In this case, it is reasonable
to assume that each loop, $a$, corresponds to a function node, and each
site, $i$ to a variable node (see figure
\ref{fig:cactusFactorGraph}). With this notation, equation (\ref{eq:hamiltonianEA})
  may be written as a sum of loop contributions $\mathcal{H}=\sum_a H_a$ where:

\begin{equation}
H_a=-\sum_i^{c+1} J_{i,i+1} S_i S_{i+1}\label{eq:functionnode}
\end{equation}

\noindent where we assumed  periodic boundary conditions in the index
$i$ within the loop. 

Then, since the resulting bipartite graph shares the structure of the Husimi lattice, it is locally tree-like and one expects the convergence of the cavity equations below.

\subsubsection{Replica Symmetric equations}

The self-consistent cavity equations in the Husimi lattice may be derived
following the approach in \cite{MP2}. Let us consider a set of $c$ cavity spins $S_1\ldots
S_c$ each of which belongs to $K$ loops and receives a field $h_i$ from
the interactions with these loops. The iteration procedure consists in
adding a new spin $S_0$ to the cavity graph together with $c+1$
couplings $J_{i,i+1}$, forming a loop (a function node) between all the $c$ cavity spins
and the spin $S_0$. With this new loop, the cavity
spins $S_1\ldots S_c$ complete their $K+1$ loops, and a field
contribution $u_{a\rightarrow 0}$, that can be computed from the
values of $h_1\ldots h_c$ and $J_{0,1}\ldots J_{c,0}$, appears on
spin $S_0$: 
\begin{eqnarray}
u_{a \rightarrow 0}=\hatu&=&\sum_{S_0} S_0 \min_{S_1\ldots S_c} H_a \nonumber \\ 
&=&-\frac{1}{2}\sum_{S_0}S_0\max_{S_1\ldots S_c}\{ J_{0,1} S_0 S_1 +\sum_{i=1}^{c-1}( J_{i,i+1} S_i S_{i+1} + h_i S_i) +h_c S_c + J_{c,0} S_c S_0\} \label{eq:hatu}
\end{eqnarray}

Then, the cavity field in the spin $S_0$ is $h_0 = \sum_{a=1  \dots K} u_a $ where all $u_a$ are computed independently by the
cavity iteration described above. The $u$ are called  cavity messages
or biases, and can
be interpreted as the contribution of a given loop $a$ to the cavity
field in $S_0$. The cavity fields $h_1, h_2, \ldots h_c$ entering a loop, will be referred from now on as
$\underline{h}$ for notation clarity. Similarly $\underline{J}$ will stand for the set of
all  the $c+1$ coupling constants, $J_{i,j}$, of the function node.
Equation (\ref{eq:hatu}) is nothing but the difference between the minimum energy
configuration of the spins at the loop when the cavity spin
$S_0$ is up and down. Following equation (\ref{eq:hatu}) it is 
easy to prove that independently of $c$ and $K$
the cavity messages in this model take only integer values: $u \in \{-2,-1,0,1,2\}$. 

Furthermore, it is expected that the distribution of these fields is
stable during the iteration procedure\cite{MP2}. This consideration yields the
self-consistent cavity equations for the fields distribution $\mathcal{P}(h)$ and the messages distribution $\mathcal{Q}(u)$:
\begin{eqnarray}
\mathcal{P}(h)=\int
\delta(h-\sum_{a} u_a)  \prod_{a=1}^{K} \ud\mathcal{Q}(u_a)\label{eq:phvsqu} \\
\mathcal{Q}(u)=\mathrm{E}_J\int  \delta(u-\hatu)\prod_{i=1}^{c} \ud\mathcal{P}(h_i)
\end{eqnarray}

\noindent $\mathcal{P}(h)$ and $\mathcal{Q}(u)$ are distributions of fields and
biases that represent the probability of finding a field or bias in any
site of the graph and $\mathrm{E}_J$ is an average over the $J_{i,j}$.
These two equations can be folded into one that is
equivalent to the ``Parent to Child'' messages-passing algorithm
described in \cite{yedidia} for single instances:

\begin{eqnarray}
\mathcal{Q}(u)=\mathrm{E}_J\int \delta(u-\hat{u}(\underline{J}, \underline{\sum^{K} u}))\prod_{i=1}^{c K} \ud\mathcal{Q}(u_i) 
\label{eq:messpas}
\end{eqnarray}

This is a self-consistent equation for the RS order parameter $\mathcal{Q}(u)$. As the biases are integers $u \in \{-2,-1,0,1,2\}$, $\mathcal{Q}(u)$
can be parameterized with five numbers $p_{-2},p_{-1},p_0,p_1,p_2$:

\begin{equation}
\mathcal{Q}(u)=p_{-2}\delta_{u,-2}+p_{-1}\delta_{u,-1}+p_{0}\delta_{u,0}+p_{1}\delta_{u,1}+p_{2}\delta_{u,2}
\label{eq:param}
\end{equation}

\noindent (or four if we consider the normalization constraint) that represent the
probabilities $p_u$ of finding a message $u$ going from a function node
to a site.

With the help of (\ref{eq:param}) the self-consistent equation
(\ref{eq:messpas}) may be written as a 
 set of five equations relating the parameters
 $p_{-2},p_{-1},p_0,p_1,p_2$. Then, the  right-hand side of
(\ref{eq:messpas}) constitutes a sum of terms of degree $c \cdot K$ in
the probabilities $p$, and degree $c+1$ in $\rho$. These equations are highly
coupled and non-linear: 
all $5^{cK}$ combinations of fields and $2^{c+1}$ combinations of
couplings have to be considered. While in some cases one is able to find
analytical solutions  for all $\rho$, in
more general situations  equation (\ref{eq:messpas}) must be solved numerically.

Once the RS solution $\mathcal{Q}_{RS}(u)$ of the self-consistent
equation (\ref{eq:messpas}) for a given $\rho$ is known, this can be used to compute the expected value of the energy in the graph, as:

\begin{equation}
U=-\frac{K+1}{c+1}\mathrm{E}_J \int
\hat{a}(\underline{J},\underline{\sum^{K}u_i})\prod^{cK}\ud \mathcal{Q}(u_i) 
+\frac{(Kc-1)}{c+1} \int |\sum u_i| \prod^{k+1}\ud \mathcal{Q}(u_i) \label{eq:RSEnergy}
\end{equation}

\noindent where the function  $\hat{a}(\cdot)$ is an energetic term very similar to the function $\hat{u}(\cdot)$:

\begin{equation}
\hata=-\frac{1}{2}\sum_{S_0}\max_{S_1\ldots S_c}\{ J_{0,1} S_0 S_1
+\sum_{i=1}^{c-1}( J_{i,i+1} S_i S_{i+1} + h_i S_i) +h_c S_c + J_{c,0}
S_c S_0\}
\label{eq:def-a}
\end{equation}

\noindent and is exactly the energy gained by the local state of the graph when the spin $S_0$ is added to the cavity in the iteration procedure.

The RS solution is valid if the system has
a single pure state. In the more general case in
which the system has many pure states it usually fails. The reason is that
within the RS approximation one assumes that under the process of
iteration the ground states of the graphs with $N$ and $N+1$
spins are related. This is not necessary true\cite{MP2}, and in such cases one must go beyond the RS solution to the 1RSB
approximation. Another, perhaps, more numerical intuition is that running
the message-passing algorithm, i.e.  equation (\ref{eq:messpas}), over
a single graph one observes that the RS
equations have multiple fixed points, which is a signature of the appearance of multiple stationary states for the system\cite{LK}.

\subsubsection{1RSB equations}
\label{sec:int-1rsb}

The 1RSB cavity method assumes that there exists an exponential number $\mathcal{N}(\epsilon) \sim exp(N \Sigma(\epsilon))$ of pure states\cite{MPV, MP2} with intensive energy $\epsilon=E^{\alpha}/N$, characterized by the complexity function $\Sigma(\epsilon)$. In this case the correct description of the interaction between a loop (node) and a spin is not given by a single message $u$, but by a distribution $Q(u)$ representing all different messages in all pure states, and the correct order parameter is now the functional $\mathcal{Q}[Q(u)]$ defining the probability of finding a distribution $Q(u)$ in a randomly selected node of the graph. The hypotheses made for the RS approximation are valid in each pure state $\alpha$, but the statistics are more subtle, since the ground state can move from one pure state to another when the iteration procedure is carried out \cite{MP1, MP2}. The self-consistent equation for the new order parameter is similar to the RS equation for $\mathcal{Q}(u)$:
\begin{eqnarray}
\mathcal{Q}[Q]=\mathrm{E}_J\int \delta^{(F)}(Q-\hat{Q}[\underline{J},\underline{Q_i}])) \prod_{a=1}^{c K} \ud\mathcal{Q}[Q_a]\label{eq:selfconsQQ}
\end{eqnarray}
\noindent where $\underline{Q_a}$ represents the set of all the $c K$
distributions $Q_a$ in the cavity iteration procedure. The functional delta accounts for the iteration of messages, and is defined by:
\begin{eqnarray}
\hat{Q} \left[ \underline{J},\underline{Q_a} \right](u) =\frac{1}{A[\underline{J},\underline{Q_a}]}\int \:e^{\mu\hatau}\delta(u-\hatuu) \:\prod_{a=1}^{c K} \ud Q_a(u_a) \label{eq:qu1RSBqu}
\end{eqnarray}

 \noindent where the exponential Boltzmann factor is the so called ``re-weighting'' that takes into account the energy shifts between states when the iteration procedure is carried out \cite{MP2}. Note that the function $\hat{a}(\cdot)$ was already defined in eq. (\ref{eq:def-a}).

 In short, in the 1RSB approximation each of the states appears with a weight proportional to $e^{- \mu f}$
 where this $\mu$ plays the role of an inverse temperature and $f$ is the corresponding free energy of the state\cite{MP1}. In the replica language the parameter $\mu$ stands for the product of the Parisi's replica
parameter $m$ and the inverse temperature $\B$ in the zero temperature
limit\cite{MP1,MP2}.  The term $A[\underline{J},\underline{Q_a}]$ is a normalization constant given by:

\begin{equation}
A[\underline{J},\underline{Q_a}]=\int e^{\mu \hatau} \prod_a^{c K}\ud Q_a(u) 
\label{eq:A}
\end{equation}

\noindent and can be thought as a local partition function. 

The order parameter $\mathcal{Q}_{1RSB}[Q]$ solution of equation (\ref{eq:selfconsQQ}) can be rarely found analytically, and furthermore, it depends on the variational parameter $\mu$. The 1RSB solution, then, is quite more involved than the RS one, not only because the order parameter becomes a functional, but also because this functional has to be found numerically, and extremized over the parameter $\mu$, as is explained next.

\subsubsection{Free energy, energy and complexity}

Within the 1RSB cavity formalism\cite{MP1,MP2}, a generalized free
energy, $\phi(\mu)$,  can be defined as :
\begin{eqnarray}
\phi(\mu)=<\phi_{site}(\mu)>-\frac{(K+1)c}{c+1}<\phi_{node}(\mu)> \label{eq:parisiphi}
\end{eqnarray}
\noindent where the $\phi_{node}(\mu)$ is the contribution from the
loops and $\phi_{site}(\mu)$ is the contribution coming from
individual sites in the Husimi graph. 
The previous expression for $\phi(\mu)$ reduces to the one in
\cite{MP2} for the EA model in the Bethe lattice, when $c=1$. The site and node contributions are given by:
\begin{eqnarray}
e^{-\mu \phi_{site}}=Z_{site}&=& \int  e^{\mu(\sum^{K+1}\hata+|\sum^{K+1}\hatu|)} \prod^{c(K+1)}\ud P_i(h_i)\nonumber \\
e^{-\mu \phi_{node}}=Z_{node}&=&\int  e^{\mu(\hata+|h_0+\hatu|)} \prod^{c+1}\ud P_i(h_i)\label{eq:Zvsph}
\end{eqnarray}

\noindent where $Z_{site}$ and $Z_{node}$ are  partition functions that assure the normalization of
the probability distributions at site and loop levels respectively.

Local distributions of fields $P(h)$ and messages $Q(u)$ hold the same relation than $\mathcal{P}(h)$ and $\mathcal{Q}(u)$ in equation  (\ref{eq:phvsqu}), though they have different meanings. Using this relation and equation (\ref{eq:A}), the computation of  $\phi(\mu)$ in (\ref{eq:parisiphi}) can be written in terms of $\mathcal{Q}[Q(u)]$ as:
\begin{equation}
-\mu\phi(\mu)=\frac{K+1}{c+1}<\log A[\underline{J},\underline{Q_a}]>-\frac{(Kc-1)}{c+1}<\log \int  e^{\mu|\sum u_i|}\prod^{K+1}\ud Q_i(u_i)> \label{eq:freeEnergy1RSB}
\end{equation}
\noindent where $<\cdot>$ stands for the average over the order
parameter distribution $\int \mathcal{Q}[Q]$ as well as over the
quenched disorder distribution $\mathrm{E}_J$. This is not, however, a
variational expression for $\phi(\mu)$ and gives only valid results
when the order parameter $\mathcal{Q}[Q]$ solves the self-consistent
equation  (\ref{eq:selfconsQQ}). The results obtained for the Replica
Symmetric case can be recovered from the 1RSB expressions by taking the $\mu\rightarrow 0$ limit. 

The complexity is related to the free energy $\phi(\mu)$ via a Legendre transform \cite{MP1, MP2}:
\begin{eqnarray}
\Sigma(\mu)=\mu \left[ \epsilon(\mu) -\phi(\mu)\right] \label{eq:complexityEphi}
\end{eqnarray}

This justifies calling $\phi$ a free energy, and $\Sigma$ and $\mu$ can be thought as the entropy and the temperature of a system whose configuration space is given only by the pure states of our original problem. On the other hand, given the relation of the complexity with the exponential abundance of states $\mathcal{N}(\epsilon) \sim exp(N \Sigma(\epsilon))$, it is clear that the ground state of the system is one with zero complexity $\Sigma=0$. This defines the point $\mu^*$ such that $\Sigma(\mu^*)=0$ and $U_{1RSB}=\epsilon(\mu^*)=\phi(\mu^*)$ is the ground state prediction at the 1RSB approximation. Furthermore, given the usual Legendre relations

\begin{eqnarray}
\epsilon(\mu)=\partial_\mu \left[\mu\phi(\mu)\right] 
&\hspace{1cm}\Sigma(\mu)=\mu^2 \partial_{\mu}\phi(\mu)  \label{eq:legendrerelations}
\end{eqnarray}

it is clear that the point $\mu^*$ extremizes $\phi(\mu)$. Following the analogy with the replica method \cite{MPV, MP2}, it can be shown that $\mu^*$ actually maximizes $\phi(\mu)$. As usual, it is sufficient to know $\phi(\mu)$ to obtain all other thermodynamic potentials, but let us mention that explicit expressions for $\epsilon(\mu)$ and $\Sigma(\mu)$ in terms of the order parameter can be derived straightforwardly \cite{MP1,MP2}.

To finish the presentation of the 1RSB cavity method some words must be said about the numerical method used to solve the equation (\ref{eq:selfconsQQ}) . 
Since $\mathcal{Q}[Q(u)]$ is a mathematical object very hard to deal with, this 
self-consistent equation is rarely solved analytically. The usual approach \cite{MP1,MP2, ricci, LK, Col2} is to represent the order parameter by a large population of distributions $Q(u)$, and to use a fixed-point method to solve eq. (\ref{eq:selfconsQQ}). In this work,  each distribution $Q(u)$ is represented by the five numbers $p_{-2}\ldots p_{2}$ (similar to (\ref{eq:param})), and the order parameter consists of a population of $N \sim 10^4$ of such distributions. For each value of $\mu$, the solution of the self-consistent equation is found by replacing several times ($\sim 10^2 N$) a randomly selected member of the distribution by the result of (\ref{eq:qu1RSBqu}). After convergence, the free energy (\ref{eq:freeEnergy1RSB}) is computed using other $\sim 5*10^3 N$ steps of the population dynamics algorithm. 

\subsection{RS-1RSB Stability Analysis}
\label{sec:sta}

The stability under 1RSB perturbations of the RS solutions as a function of $\rho$  can be studied with the method applied in \cite{ricci, ricciMon}. The replica symmetric self-consistent equation can be recovered from the 1RSB \cite{ricci}, by restricting the space of the 1RSB order parameter:

\begin{equation}
\mathcal{Q}[Q]\rightarrow \mathcal{Q}_{RS}[Q]=\sum_{q=-2}^{2}p_{q} \delta^F(Q(u)-\delta_{u,q}) \label{eq:RSfrom1RSB}
\end{equation}

In such a case,  each distribution is forced to be deltaic and the re-weighting terms in (\ref{eq:qu1RSBqu}) can be factorized out of the integrals. Then, all 1RSB expressions turn to be exactly the ones obtained at the RS level.

It is now fruitful to study the effect of the iteration procedure over a 1RSB parameter that is almost RS, except for a small perturbation of non-deltaic messages distributions:
\begin{eqnarray}
\mathcal{Q}[Q]&=& \sum_{q=-2}^{2}p_{q} \delta^F(Q(u)-\delta_{u,q})\label{eq:generalPerturb} \\
&&+\sum_{n=1}^{26} \int \pi_n(\{y\}) \delta^F(Q(u)-Q_n(\{y\})(u)) \ud \{y\}\nonumber
\end{eqnarray}
\noindent and to see how it evolves. The symbol $\{y\}$ stands for a subset of the parameters $y_2,y_1,y_{-1},y_{-2}$ (see equation below). Following \cite{ricci, ricciMon} this means that the order parameter has the following composition:

\begin{equation}
Q(u)=\left\lbrace  
\begin{array}{ll}
\delta_{u,i} & \mbox{with probability    } p_{i}, i\in(-2\ldots2) \\ 
y_2 \delta_{u,2}+(1-y_2) \delta_{u,1}&\mbox{with probability    } \pi_1(y_2) \\
y_2 \delta_{u,2}+(1-y_2) \delta_{u,0}&\mbox{with probability    } \pi_2(y_2) \\
\hspace{1cm}\vdots &\vdots \\ 
y_2 \delta_{u,2}+y_1 \delta_{u,1}+ (1-y_1-y_2) \delta_{u,0}& \mbox{with
  probability    } \pi_{11}(y_1,y_2)\\
\hspace{1cm}\vdots &\vdots \\
y_2 \delta_{u,2}+y_1 \delta_{u,1}+& \\
(1-y_1-y_2-y_{-1}-y_{-2})\delta_{u,0}+& \mbox{with probability    } \pi_{26}(y_{-2},y_{-1},y_{1},y_{2}) \\
y_{-1} \delta_{u,-1}+y_{-2} \delta_{u,-2} & \\
\end{array}\right.
\label{eq:perturb}
\end{equation}
\noindent where the perturbation part is composed of all the $26$
possible combinations of non-delta shaped distributions. We will refer to them as $Q_n(\{y\})(u)$
in general, and sometimes dropping the $u$ for clarity. $Q_n(\{y\})(u)$ shall be interpreted as a function of $u$, where $n$ and   $(\{y\})$ are parameters, the former qualifying the type of non-deltaic distribution, and the latter defining the probabilities of each message in the distribution. Let us highlight that, while 
 $Q_{26}(\{y\})(u)$ seems to contain all the other possibles $Q_{n}(\{y\})(u)$, it is not the case, since, to be meaningful in the context of the decomposition (\ref{eq:perturb}), all the components of the vector $\{y\}$ must be greater than zero.  This is a way to focus our attention directly on the overall weight of each type of perturbation:
\begin{equation}
\Pi_n=\int \pi_n(\{y\})\ud \{y\} \label{eq:overallEa}
\end{equation}

For a first order study of the stability, the $\pi_n(\{y\})$ are supposed to be infinitesimal quantities such that, during the iteration, the presence of more than one non-delta
shaped distributions in the cavity will be very improbable. The convolution of a
given $Q_{n}(\{y\})$ with $c K-1$ delta-shaped distributions in
eq. (\ref{eq:qu1RSBqu}) can yield, depending on the realization
of the coupling constant, any of the
distributions in (\ref{eq:perturb}). Considering all the possibilities and
their probability of occurrence in eq. (\ref{eq:qu1RSBqu}), 
we can construct the matrix $\mathbf{I}_{26\times26}$ of elements
$i_{n,m}$ giving the probability that a non-delta-shaped distribution
of type $n$ turns into type $m$ after the iteration. Then, after the
iteration, the vector $\overrightarrow{\Pi}=(\Pi_1,\Pi_2,\ldots,\Pi_{26})$ of the overall weight of the perturbations is transformed by
\begin{equation}
\overrightarrow{\Pi'}=\mathbf{I}\overrightarrow{\Pi} \label{eq:vectorweightevolution}
\end{equation}

A given RS solution is stable if all the eigen-values of $\mathbf{I}$ are smaller than one:
\begin{equation}
\max \lambda (\mathbf{I})<1
\end{equation}

This is the general approach in studying the RS stability. In some cases, this
approach may lead to simple and closed equations that allow analytical
solutions, but in
general,  the properties of the matrix  $\mathbf{I}$ must be studied numerically.

\section{Results for the triangular Husimi lattice}
\label{sec:triang}

In this section we apply the formalism described above to solve, at $T=0$, the Edwards-Anderson model in the
simplest of all Husimi lattices: the triangular lattice, with $c=2, K=1$
 (see fig \ref{fig:cactusFactorGraph}). Note that the case, $\rho=-1$ was recently extensively studied in \cite{luca} and is equivalent to the bi-coloring problem in an hyper-graph. Here we will keep  $K=1$ and concentrate our attention on the role of $\rho$ in the thermodynamics of the system.

\subsection{RS-solution}\label{sec:RSsolution}

This graph
allows closed expression for all the RS fixed points. For example, a careful analysis of 
(\ref{eq:messpas}) and (\ref{eq:param}) for this lattice leads to the
definition of the following variables: $m_1=p_{1}-p_{-1}$,
$s_1=p_{1}+p_{-1}$, $m_2=p_{2}-p_{-2}$ and
 $s_2=p_{2}+p_{-2}$, such that (\ref{eq:messpas}) may be written as:
\begin{eqnarray}
m_1&=&(m_1p_0+\frac{s_1}{2}(m_1+2 m_2))\rho-m_1(\frac{s_1}{2}+s_2-p_0)\rho^2 \nonumber \\
s_1&=&s_1(p_0+s_2+\frac{s_1}{4})-m_1(m_2-\frac{m1}{4})\rho+\frac{m_1^2}{4}\rho^2+(p_0-\frac{s_1}{4})s_1\rho^3\nonumber \\
m_2&=& (s_1m_1+s_2m_1-s_1m_2+2 m_2)\frac{\rho}{2}+(m_1s_1+s_2m_1+s_1m_2+2 p_0m_2)\frac{\rho^2}{2} \nonumber\\
2 p_0&=&s_1^2+s_2^2+s_1s_2+2 p_0+m_1m_2\rho-(m_1^2+m_2^2+m_1m_2)\rho^2-(2 p_0(1-p_0)+s_1s_2)\rho^3\nonumber \\
s_2&=&1-p_0-s_1 \label{eq:cactusEqsm}
\end{eqnarray}

It is easy to check that, independently of $\rho$, a trivial
paramagnetic solution (P) where only $u=0$ messages have a non-zero
probability $p_0=1$, always exists:

\begin{equation}
\mathcal{Q}_{P}(u)=\delta_{u,0}
\end{equation}

\begin{figure}[!htb]
\begin{center}              
            \includegraphics[scale=0.65,angle=0]{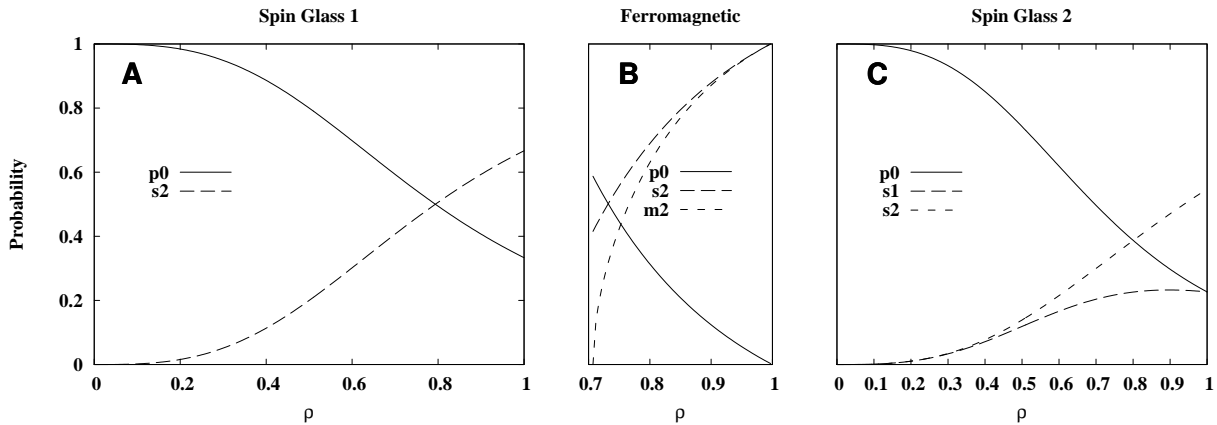} 
           \end{center}
	\caption{The solution of the RS equations for the triangular Husimi lattice ($c=2$ and
          $K=1$). Panel {\bf A} is the first spin-glass solution
          (SG-1). Panel {\bf B} represents the ferromagnetic
          solution. In panel {\bf C} appears the second spin-glass
          solution (SG-2). The variables are defined in section \ref{sec:RSsolution}.}
	\label{fig:RSsolut}
\end{figure}

On the other hand, setting the values of $m_1$ and $s_1$ to zero, i.e. $p_1=p_{-1}=0$, automatically satisfies
the first two equations in (\ref{eq:cactusEqsm}). With the three
remaining equations, we can find three solutions. The first one is a spin glass
paramagnetic solution (SG-1) with $p_0=(2\rho^3+1)^{-1}, m_2=0, s_2=1-p_0$: 

\begin{eqnarray}
\mathcal{Q}_{SG-1}(u)=\frac{\rho^3}{2\rho^3+1}\delta_{u,-2}+\frac{1}{2\rho^3+1}\delta_{u,0}+\frac{\rho^3}{2\rho^3+1}\delta_{u,2} \label{eq:RSpara1SoluRho}
\end{eqnarray}

\noindent that is valid for $\rho\geq\rho_{SG-1}=0$ and is shown in panel {\bf A} of figure \ref{fig:RSsolut}. The second and third solutions are actually the same 2-degenerated ferromagnetic solution with $p_0=\frac{1-\rho}{\rho^2}$, $s_2=1-p_0$ and $m_2^2=(\rho^2+\rho-1)^2\rho^{-6}-2(\rho^2+\rho-1)(1-\rho)\rho^{-3}$:

\begin{equation}
\mathcal{Q}_{F}(u)=\frac{s2-m2}{2}\delta_{u,-2}+\frac{1-\rho}{\rho^2}\delta_{u,0}+\frac{s2+m2}{2}\delta_{u,2} \label{eq:RSferroSoluRho}
\end{equation}

\noindent valid for $\rho\geq \rho_{F}=\frac{1}{\sqrt{2}}\simeq
0.7071$. The dependencies of $p_0$, $s_2$ and $m_{-2}$ with $\rho$ appear in panel {\bf B} of figure \ref{fig:RSsolut}.

There is yet another spin glass paramagnetic solution with $s_1\neq0$,
$m_1=0$ and $m_2=0$ that we will call SG-2. It differs from the SG-1 in that $p_1=p_{-1}>0$. The
dependencies of $s_1$, $s_2$ and $p_0$ for this solution are shown 
in the panel {\bf C} of figure \ref{fig:RSsolut}. Like the other
solutions described above, this solution can be analytical expressed as a function of
$\rho$. However, the resulting algebraic expression is quite large and
does not add clarity to the remaining discussion, so we do not write it
explicitly here. Just note (figure \ref{fig:RSsolut}) that this solution exists only for  $\rho\geq\rho_{SG-2}=0$. At this point it is worth mentioning that the notations SG-1 and SG-2 refer to the non-trivial structure (although symmetric) of the distributions, but one must keep in mind that 
 the solution is ergodic and does not represent a true glassy system.

\begin{figure}[!htb]
\begin{center}              
            \includegraphics[scale=0.8,angle=0]{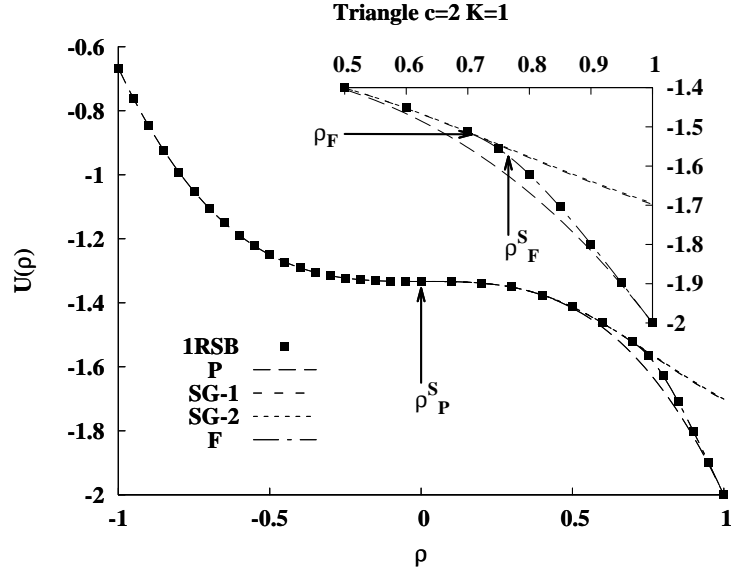} 
           \end{center}
	\caption{Energies for the different RS solutions
          and the 1RSB approximation $\phi(\mu^*)$ as a function of $\rho$ for the triangular Husimi
          lattice ($c=2$, $K=1$). We represent also the stability
          threshold $\rho_P^S=0$ of the trivial paramagnetic solution. The caption shows the point in
          which the ferromagnetic solution appears $\rho_F$, and the point where it becomes
          stable $\rho_F^S$.}.
	\label{fig:RSEnergy}
\end{figure}

Each of the four solutions described above defines an energy function $U(\rho)$ by eq. \ref{eq:RSEnergy}) (see figure \ref{fig:RSEnergy}). Except for the second spin glass paramagnetic solution, which we leave to the interested reader to work out, the expressions for the energy are rather simple:

\begin{eqnarray}
U_{P}&=&-\frac{2}{3}(2+\rho^3) \\ \nonumber
U_{SG-1}&=&-\frac{2}{3} \frac{ 2 + 9\rho^3 + 12\rho^6}{(1 +
  2\rho^3)^2} \\ \nonumber
U_{F}&=&-\frac{2}{3} \frac{(-1 + 3\rho - 7\rho^3 + 6\rho^4 - \rho^6 + 3\rho^7)}{3\rho^6}
\label{eqn:ERS}
\end{eqnarray}

Summarizing, the RS calculations show that: 
{\it i)} a trivial paramagnetic solution exists for all $\rho$; {\it ii)} at $\rho=\rho_{SG-1}=\rho_{SG-2}=0$ two spin glass
paramagnetic solutions appear and are valid up to
$\rho=1$; {\it iii)} for $\rho>\rho_{F}=2^{-1/2}$ one also finds 
a ferromagnetic solution. Note that, in the scale of the figure
\ref{fig:RSEnergy}, it is hard to differentiate the actual energies of
the different solutions for a wide range of
values of $\rho$. As we demonstrate in the next section, the same is true when
compared with the 1RSB solution. However, we will show that for
other lattices the situation is different.

\subsection{Stability analysis of the RS solutions}

Though the general method given in the introduction is always valid,
we will try to keep 
 the stability analysis as simple as possible. The trivial
 paramagnetic solution can be studied, for instance, with only one
 general perturbation 
$Q_{26}(y_{-2},y_{-1},y_{1},y_{2})= y_{-2} \delta_{u,-2}+y_{-1} \delta_{u,-1}+(1-y_{-2}-y_{-1}-y_{1}-y_{2}) \delta_{u,0}+y_{1} \delta_{u,1}+y_{2} \delta_{u,2}$

\begin{eqnarray}
\mathcal{Q}[Q]&=& \delta^F(Q(u)-\delta(u,0)) \label{eq:paramPerturb} \\
&&+\int \pi(y_{2},y_{1},y_{-1},y_{-2}) \delta^F(Q(u)-Q_{26}(y_{2},y_{1},y_{-1},y_{-2})) dy_2 dy_1 \ldots \nonumber
\end{eqnarray}

The idea is to follow the overall weight of the perturbation $\Pi=\int
\pi(y_{2},y_{1},y_{-1},y_{-2})\ud y_{2}\ud y_{1}\ud y_{-1}\ud y_{-2}$ for
the iteration  (\ref{eq:qu1RSBqu}). The reason behind
the success
of such a simple analysis  lies in the fact that the iteration
eq. (\ref{eq:qu1RSBqu}) has a very naive behavior when we
consider the convolution of the paramagnetic solution with a
perturbation. Depending on the
realization of the  coupling constants $J_{i,j}$, the convolution 
of the perturbation $Q_{26}(y_{-2},y_{-1},y_{1},y_{2})$ and the RS solution
$\delta(u,0)$ can only produce one of the following three results:

\begin{eqnarray}
\delta_{u,0}\ast Q_{26}(y_{-2},y_{-1},y_{1},y_{2}) &\rightarrow& \delta_{u,0} \nonumber \\
\delta_{u,0}\ast Q_{26}(y_{-2},y_{-1},y_{1},y_{2}) &\rightarrow& Q_{26}(y_{2},y_{1},y_{-1},y_{-2}) \nonumber \\
\delta_{u,0}\ast Q_{26}(y_{-2},y_{-1},y_{1},y_{2}) &\rightarrow& Q_{26}(y_{-2},y_{-1},y_{1},y_{2}) \nonumber 
\end{eqnarray}

\noindent where we are using the symbol $\ast$ to represent the right-hand side of
eq. (\ref{eq:qu1RSBqu}). Averaging over all possible couplings,
we found that the last two situations take place with a probability
$\frac{1+\rho^3}{2}$. Then, also taking into account a combinatorial
factor of 2, the overall weight of the perturbation evolves following:

\begin{equation}
\Pi'_{P}=(1+\rho^3)\Pi_{P}
\end{equation}

\noindent showing that for all $-1 \leq\rho< \rho_{P}^{S}=0$ the
paramagnetic solution $Q_{P}(u)=\delta_{u,0}$ is stable, while it is unstable for
$\rho>0$.  To determine the stability of the solution  at $\rho_{P}^{S}=0$ one must use
a second order perturbation.

Unfortunately, in general, such a simple analysis is not applicable. 
For instance, the spin glass paramagnetic solution, SG-1, when
convolved with $Q_{26}$, can yield mixed
distributions of other types. Then, in order to keep
 analytical expressions, we
consider in this case two  other types of perturbations:
\begin{eqnarray}
Q_n(y)&=& (1-y)\delta_{u,0}+y\delta_{u,2} \nonumber \\
Q_m(y)&=& y\delta_{u,-2}+(1-y)\delta_{u,0}  
\end{eqnarray}

Both the spin glass paramagnetic solution, SG-1, and the ferromagnetic
solution, F, have only messages $ u\in\{-2,0,2\}$.  Then, the convolution with the perturbations $Q_{n,m}$ produces one of the following results:
\begin{eqnarray}
Q_{SG-1,F}\ast Q_{n,m}(y)  &\rightarrow& \delta_{u,0} \nonumber \\
Q_{SG-1,F}\ast Q_{n,m}(y)  &\rightarrow& \delta_{u,2} \nonumber \\
Q_{SG-1,F}\ast Q_{n,m}(y)  &\rightarrow& \delta_{u,-2} \nonumber \\
Q_{SG-1,F}\ast Q_{n,m}(y)  &\rightarrow& Q_{m}(y') \nonumber \\
Q_{SG-1,F}\ast Q_{n,m}(y)  &\rightarrow& Q_{n}(y') \nonumber 
\end{eqnarray}

\noindent where only the last two situations are responsible for the
evolution of the perturbation. This set of perturbations is
closed under the iteration equation (\ref{eq:qu1RSBqu})
and therefore suitable for analytical treatment.

For the spin glass solution, SG-1, the matrix for the evolution of the perturbation $\mathbf{I}_{2\times 2}^{SG-1}$ is symmetric, with elements:

\begin{eqnarray}
i_{n,n}=i_{m,m}=\frac{1+3\rho^3+\rho+\rho^2-2\rho^5}{2(1+2\rho^2)} \nonumber \\
i_{n,m}=i_{m,n}=\frac{1+3\rho^3-(\rho+\rho^2-2\rho^5)}{2(1+2\rho^2)} \nonumber 
\end{eqnarray}

The largest of its eigenvalues is $\lambda=\frac{1+3\rho^3}{1+2\rho^3}$ which is always greater than one, for $\rho>0$. This means that the spin glass SG-1 solution is always unstable.

For the ferromagnetic solution, the matrix  $\mathbf{I}_{2\times 2}^{F}$ has the following structure

\begin{eqnarray}
i_{n,n}=i_{m,m}=\frac{1}{2}(1+\frac{1}{\rho}-\rho-2\rho^2) \nonumber \\
i_{n,m}^+,i_{m,n}^-=\frac{3\rho^2-1}{2\rho}\pm \frac{1}{\rho^2}\sqrt{(\rho^4-(\rho-1)^2)(2\rho^2-1)}\nonumber 
\end{eqnarray}

\noindent where the terms $i_{n,m}^+,i^-_{m,n}$  differ in the sign in
front of  the square root. The limiting condition for the stability $\max
\lambda (\mathbf{I})=1$ turns out to be equivalent to the solution of
$\rho^6+\rho^4-2\rho^3+\rho^2+\rho-1=0$. This defines the point
$\rho_{F}^{S}\simeq0.765942$ where the ferromagnetic solution becomes
stable. 

For the spin glass paramagnetic solution SG-2 it was impossible to find a
closed  subset of perturbations smaller than the full set. Therefore we studied its
stability using the full matrix $\mathbf{I}_{26\times 26}$ . Given the
size of the matrix, and the non-simple dependence of this solution
with $\rho$, the study of the stability was done numerically. We calculated
all the 26 eigenvalues of $\mathbf{I}_{26\times 26}$ and found that some
of them were always above 1 in the interval $0<\rho<1$. This means that
the spin glass solution (SG-2), like  the (SG-1), is always unstable.

Note, that the results obtained for the SG-1 and ferromagnetic solutions
where only a subset of perturbations were studied, can be considered only as
lower-bound approximations for the stability. For the SG-1 the
full $26$ perturbations procedure can be done analytically, and it
appears that the highest eigenvalue was, in fact, the one we found with
the restricted perturbation method. 
For the ferromagnetic solution it is harder to get an analytic
expression for the eigenvalues of the 
full matrix $\mathbf{I}_{26\times 26}$, but yet we checked numerically
that all the eigenvalues  of this matrix were indeed equal or smaller than the one found within the restricted perturbation analysis.
In summary, the triangular Husimi lattice with $K=1$ has a stable
RS trivial paramagnetic solution for
$-1\leq\rho<\rho_{P}^{S}=0$, has a stable ferromagnetic
solution $Q_{F}(u)$ when $\rho_{F}^{S}\leq\rho<1$, and
two spin glass paramagnetic solutions that are unstable to
replica symmetry breaking. These points of stability are shown in
figure \ref{fig:RSEnergy}.

\subsection{1RSB solution}

The 1RSB approximation for the free energy of a model is a 
formalism more general than the Replica Symmetric calculation. Thus,
in the calculation of the free energy, we do not need to restrict ourselves to the interval $\rho^{S}_{P}<\rho<\rho_{F}^{S}$ where the RS solutions are unstable. Instead we can do the 1RSB calculation in the whole interval $-1\leq\rho\leq1$, and the RS solutions will still be found if they are thermodynamically significant, which is a stronger condition than stability.

Using the population dynamics described in section \ref{sec:int-1rsb} we computed $\phi(\mu)$ for different $\mu$.  In figure \ref{fig:RSEnergy} we compare the ground-state predictions $U_{RS}$ and $U_{1RSB}$ for both approximations (RS and 1RSB) as a function of $\rho$. In the caption we show how the 1RSB solution
collapses into the F solution for $\rho>\rho_F^S$. Unfortunately,
with the resolution of figure \ref{fig:RSEnergy} it is impossible to
differentiate the ground state energy in the 1RSB approximation from
the energy of the RS solutions within the region in which no RS solution is stable, although the order parameter $\mathcal{Q}[Q]$ showed to be non-trivial, i.e. not having the form of (\ref{eq:RSfrom1RSB}).
\begin{figure}[!htb]
\begin{center}              
            \includegraphics[scale=0.8,angle=0]{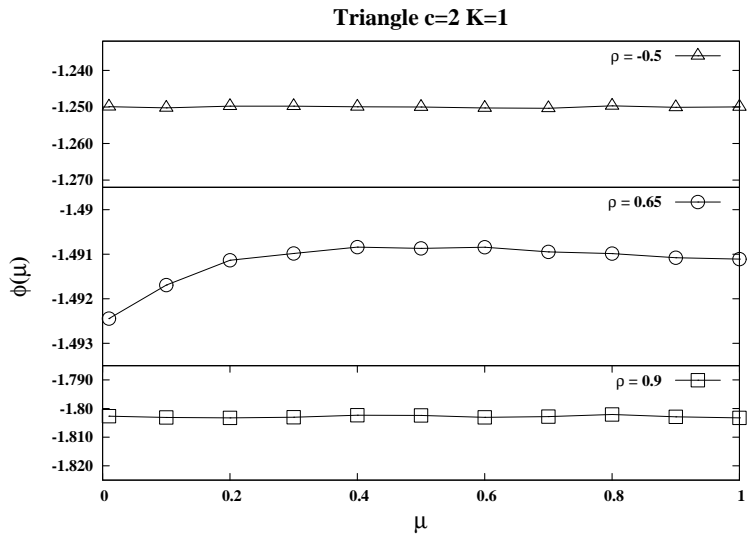} 
           \end{center}
	\caption{Free energy $\phi(\mu)$ of the 1RSB solution for the triangular Husimi lattice
        ($c=2$, $K=1$) for different values of $\rho$. Top
        panel: paramagnetic region, Middle: 1RSB region and Bottom:
        ferromagnetic. The flat curves are representative of RS solutions, while the curve in the 1RSB region has a clear non trivial maximum $\phi(\mu^*)=U_{1RSB}$ at $\mu\simeq 0.6$.}
	\label{fig:1RSsolut}
\end{figure}

For the sake of understanding the qualitative differences between the RS and
the 1RSB solution, we show the curve $\phi(\mu)$ for three different $\rho$
in figure \ref{fig:1RSsolut}.
The lower curve corresponds to the ferromagnetic
region ($\rho > \frac{1}{\sqrt{2}}$), the upper to the 
  paramagnetic (P) interval, and the middle curve to the 1RSB
  zone. In the paramagnetic and ferromagnetic regions, $\phi(\mu)$ is flat, as expected. It has
  zero derivative and hence zero complexity, meaning that the solution space is not clustered and that the RS assumptions are valid. Furthermore, a detailed analysis of the order parameter shows that it actually has the RS structure (\ref{eq:RSfrom1RSB}). On the other hand, in the 1RSB region ($\rho =0.65$) $\phi(\mu)$ is convex and has a maximum at $\mu=\mu^*\simeq 0.6$, and the order parameter is genuine 1RSB (not made of deltaic distributions) for all values of $\mu$. The highest point of this curve defines the 1RSB approximation for the ground state energy $U_{1RSB}=\phi(\mu^*)$.

The learned reader will find rare that a highly-frustrated system, as the Husimi triangular lattice with an excess of antiferromagnetic interactions ($\rho<0$), does not have a 1RSB solution. He might argue that the proved stability of the trivial paramagnetic RS solution is not enough to rule out other solutions. In fact, the absence of 1RSB solution may well be an artifact due to the random initial conditions used in the population dynamics. Fortunately, with the same tools used to analyze the stability of RS solutions, it is possible to prove (see appendix A) that any $\mathcal{Q}[Q(u)]$ solution of the self-consistent equation (\ref{eq:selfconsQQ}) is unstable to a paramagnetic perturbation $\mathcal{Q}[Q]=\pi \delta^F(Q(u)-\delta_{u,0})$. This rules out any possible stable solution except for the complete paramagnetic one $\mathcal{Q}_P[Q(u)]=\delta^F(Q(u)-\delta_{u,0})$, rendering the trivial paramagnetic solution the only thermodynamically relevant in the region $-1\leq\rho<0$. It must be kept in mind, however, that at finite temperature a 1RSB solution  with real-valued fields, which are not embedded in our parameterization, may facilitate a RS-1RSB transition in this zone.

To help the intuition, let us look closer to the case $\rho=-1$. In this case, each triangle has a minimum energy 6-times degenerated, and only 2 excited states. It is true that in all the states of minimum energies, one of the bonds of the triangle is violated, but one cannot forget that this is anyway the ground state of the triangle. Then, because this state is highly-degenerated, it is not hard to imagine that there is a lot of freedom, even on the large scale of the network, to have all the triangles ''satisfied''. Of course, this argument is valid provided $K$ is low. If $K$ is high enough, each spin will be connected to a large number of triangles, and therefore, the freedom of choice will be reduced and eventually may disappear. This, in fact, should turn the problem to the non-trivial 1RSB scenario predicted in\cite{luca} for $K>6$.

\section{General Husimi lattices}
\label{sec:gene}

The procedure described in the introduction, and applied to the study
of the triangular Husimi graph, is also applicable to all other Husimi
graphs. The main difference arises in the difficulty in generalizing
the analytical results obtained above. For instance, the equation
(\ref{eq:messpas}) is still parameterized by
$p_{-2},p_{-1},p_0,p_1,p_2$, but since each term has degree $c \cdot K$
 on these probabilities, it is very hard to write closed 
 equations in terms of $\rho$. The absence of closed expressions  for these probabilities
 also makes it impossible to express the RS-1RSB instability analytically.
 However we can always solve numerically the set of five equations
 (\ref{eq:messpas}) on the probabilities $p_i$ for each value of
 $\rho$, and check numerically the eigenvalues of the stability matrix
 for such solutions. The only exception that allows a full
 analytical treatment independently of $K$ and $c$, 
is the trivial paramagnetic solution $Q_{P}=\delta_{u,0}$, that is
obviously a solution of equation (\ref{eq:messpas}) for all the Husimi lattices.

\subsection{The trivial paramagnetic solution}

The result obtained for the trivial paramagnetic solution in the
cactus can be easily generalized to all kinds of pure and regular
Husimi lattices, with generic loop size $c+1$ and degree $K+1$. The
paramagnetic solution is always valid, but it changes from one type of graph to another. 
To study the transition RS-1RSB we consider again a perturbed order
parameter of type (\ref{eq:paramPerturb}). In the general case equation
(\ref{eq:qu1RSBqu}) is a
convolution of $c\cdot K$ distributions. Again, the convolution of paramagnetic distributions reproduces a paramagnetic distribution:
\[
\left(\delta_{0,u} \ast \right)^{c K} \rightarrow \delta_{0,u}
\]

The non-trivial case is the convolution with the perturbed distributions $Q_{26}$ that are present in the order parameter $\mathcal{Q}[Q]$ in an amount proportional to $\pi_{26}(y_{2},y_{1},y_{-1},y_{-2})$ (see equation
 (\ref{eq:paramPerturb})) . Depending on the realization of the coupling constants, the integration of eq. (\ref{eq:qu1RSBqu}) may have three different results:
\begin{eqnarray}
\left(\delta_{u,0}\ast\right)^{cK-1} Q_{26}(y_{2},y_{1},y_{-1},y_{-2}) &\rightarrow& \delta_{u,0} \nonumber \\
\left(\delta_{u,0}\ast\right)^{cK-1} Q_{26}(y_{2},y_{1},y_{-1},y_{-2}) &\rightarrow& Q_{26}(y_{2},y_{1},y_{-1},y_{-2}) \nonumber \\
\left(\delta_{u,0}\ast\right)^{cK-1} Q_{26}(y_{2},y_{1},y_{-1},y_{-2}) &\rightarrow& Q_{26}(y_{-2},y_{-1},y_{1},y_{2}) \label{eq:ParaConvCases} 
\end{eqnarray}

It is easy to check that, independently of the loop size $c+1$ and the
connectivity $K+1$ of the Husimi graph, the last two outcomes (those
that are responsible for the propagation of the perturbation) take
place only when an even number of couplings $J_{i,j}$ are
anti-ferromagnetic. The probability to have an even number of anti-ferromagnetic couplings around a loop can be computed as:
\begin{equation}
\sum_{q=0}^{(c+1)/2} \left( \frac{1+\rho}{2}\right)^{c+1-2q}\left( \frac{1-\rho}{2}\right)^{2q}\left(\begin{array}{l}
\phantom{c}2q \\c+1
\end{array}
\right)=\frac{1}{2}(1+\rho^{c+1})
\label{eq:pr-af}
\end{equation}

The upper limit of the sum is taken to be the higher integer value not
greater than $(c+1)/2$.  Since the perturbation $Q_{26}$
can occur in any of the $c \cdot K$ cavity biases distribution of
eq. (\ref{eq:qu1RSBqu}),one needs to consider a $c K$ combinatorial factor in the
previous expression.  Thus we get that the perturbation weight
$\Pi_P=\int \pi(y_{2},y_{1},y_{-1},y_{-2})$  evolves through:
\begin{equation}
\Pi_P=\frac{c K}{2}(1+\rho^{c+1}) \Pi_P
\end{equation}

The condition of stability is the damping of the perturbation $\frac{c K}{2}(1+\rho^{c+1})<1$ which has no solution for odd $c$, i.e. for even loop sizes as squares, hexagons, etc. The general solution:
\begin{equation}
\rho<\rho_{P}^{S}=-\left(\frac{c K-2}{c K}\right)^{\frac{1}{c+1}} \label{eq:rhoparam}
\end{equation}

\noindent is valid for odd loop sizes like the triangle, the pentagon,
etc. For values of $\rho<\rho_{P}^{S}$ the paramagnetic solution
$\mathcal{Q}(u)=\delta_{0,u}$ is stable. On the other hand, for even loop sizes the
paramagnetic solution is never stable. 

The energy of the paramagnetic solution can be computed very
easily. Given that only $u=0$ messages exist, each cavity loop
preserves the symmetry of the Hamiltonian ($S\leftrightarrow -S$), and
then the only contribution to the energy comes from the frustration
inside the loops. Frustrated loops are those with an odd number of
anti-ferromagnetic interactions, and  have an energy  $-(c-1)$, while
loops with an even number of anti-ferromagnetic interactions are
 not frustrated  and have energy $-(c+1)$. Then, using (\ref{eq:pr-af})
we get  as the energy of the paramagnetic solution:
\begin{equation}
U_{P}=-\frac{K+1}{c+1}(c+\rho^{c+1}) \label{eq:generalUpara}
\end{equation}

So, two key difference arise between Husimi lattices with even and odd loop sizes. The parity of the energy function $U_{P}(\rho)$ is exactly that of the loop sizes $c+1$, \textit{i.e.} graphs with even loop sizes have an even energy function $U_{P}(\rho)=U_{P}(-\rho)$, while graphs with odd loop sizes have an odd energy dependence with $\rho$ (except for a constant), as shown in figures \ref{fig:otherHusimiUvsrho2} and \ref{fig:otherHusimiUvsrho}. Furthermore, the trivial paramagnetic solution is stable (under 1RSB perturbations) only in lattices with odd loop sizes.

\subsection{Spin glass and ferromagnetic solutions}

\begin{figure}
\begin{center}              
            \includegraphics[scale=1.0,angle=0]{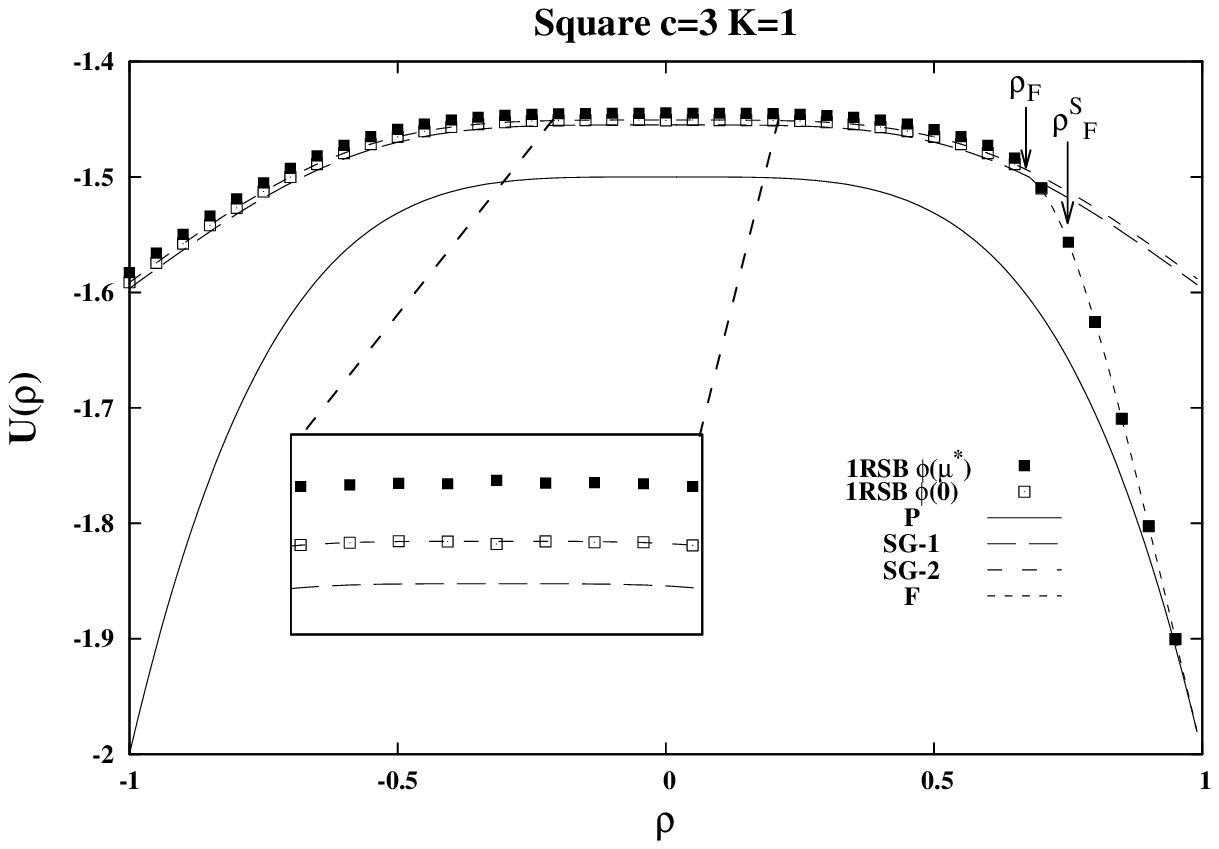} 
           \end{center}
	\caption{Energies for the different RS solutions
          and the 1RSB solution as a function of $\rho$ for the square Husimi
          lattice ($c=3$, $K=1$). The appearance and the stability
          of the solution F are \textbf{signaled}. The caption
          shows the difference between the 1RSB approximation $U_{1RSB}=\phi(\mu^*)$
          (black squares) and the energy of the RS solutions. White squares represent
          $\phi(0)$ and coincide with the energy of the SG-2  solution}
	\label{fig:otherHusimiUvsrho2}
\end{figure}

\begin{figure}
\begin{center}              
            \includegraphics[scale=1.0,angle=0]{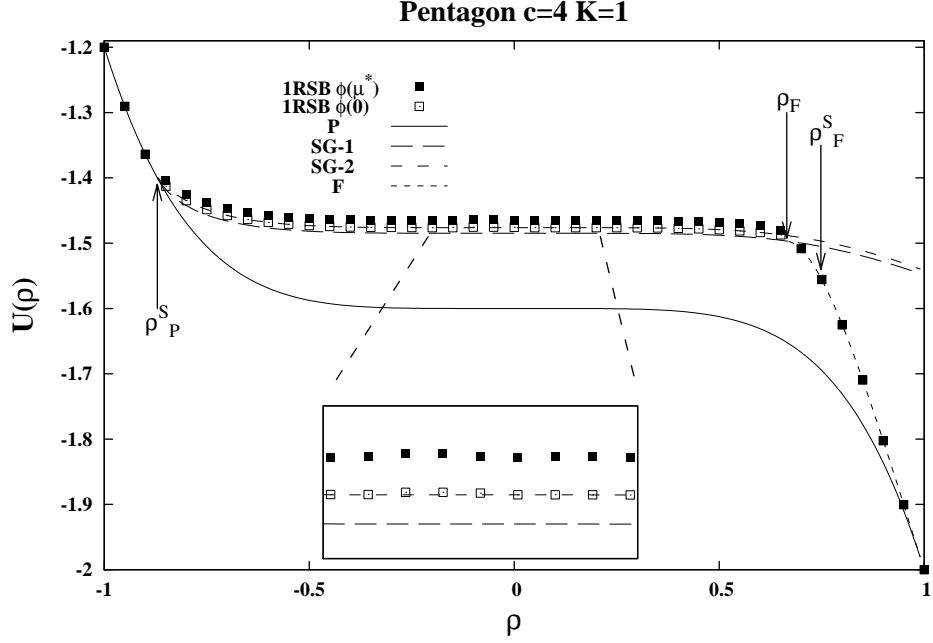} 
           \end{center}
	\caption{Energies for the different RS solutions
          and the 1RSB solution as a function of $\rho$ for pentagonal Husimi
          lattice ($c=4$, $K=1$). The stability threshold for the P and F solutions are signaled. The caption
          shows difference between the 1RSB approximation $U_{1RSB}=\phi(\mu^*)$ (black squares) and the energy of the RS solutions. White squares represent $\phi(0)$ and coincide with the energy of the SG-2  solution.}
	\label{fig:otherHusimiUvsrho}
\end{figure}

The numerical solutions of (\ref{eq:messpas}) for different values of
$c$ and $K$ show a zoology of phases similar to the one found in the triangular lattice: a trivial paramagnetic solution (P) already discussed in detail, two spin glass paramagnetic solutions (SG-1) and (SG-2), and a ferromagnetic solution (F) (see figures \ref{fig:otherHusimiUvsrho2} and \ref{fig:otherHusimiUvsrho}). The solutions SG-1 and F give zero probability to messages of type $u=\pm1$, as in the triangular Husimi lattice. 

As expected, for a sufficiently ferromagnetic systems, $\rho>\rho_{F}$, there is a ferromagnetic phase in all types of Husimi graphs. The spin glass solutions, however, are related to the parity of the loops. Husimi lattices with even loop sizes (square  $c+1=4$,  hexagon $c+1=6$, etc ) present the spin glass solutions for all $\rho$ (see figure \ref{fig:otherHusimiUvsrho2}). On the contrary, in odd loop sized Husimi graphs, the appearance of the spin glass solutions seems to be related to the loss of stability of the trivial paramagnetic solution (see figure \ref{fig:otherHusimiUvsrho}). Other Husimi lattices with odd and even number of loop sizes $c+1$ and $K=1,2$ reproduced the same behavior. 

The stability of these solutions was analyzed numerically by studying the eigenvalues
of the  $\mathbf{I}_{26\times 26}$ matrix (see section \ref{sec:sta}). In general the panorama is
similar to the one found for the triangular case;  the two spin glass paramagnetic solutions
are always  unstable to 1RSB perturbations, and the ferromagnetic
solution has a non-trivial stability point
$\rho_{F}^{S}$ above which it becomes stable (see figures \ref{fig:otherHusimiUvsrho2} and \ref{fig:otherHusimiUvsrho}). In table
(\ref{tab:values}) we present, for different  $c$ and $K$, the numerical
values for these stability points and for the appearance of the ferromagnetic solution $\rho_{F}$.

\begin{table}[!htb]
\begin{tabular}{|l @{\hspace{1cm}}|c @{\hspace{5mm}}|c @{\hspace{5mm}}|c @{\hspace{5mm}} |}
\hline Husimi      &$\rho_{P}^{S}$& $\rho_{F}$ & $\rho_{F}^{S}$ \\ 
\hline$c=2$, $K=1$ & 0 &$1/\sqrt{2}\simeq0.707 $& $0.766$ \\ 
\hline$c=3$, $K=1$ & ---- &$0.671 $ & $0.749$ \\ 
\hline$c=4$, $K=1$ & $-\sqrt[5]{\frac{1}{2}}\simeq -0.8706$ &$0.665$ & $0.748$ \\
\hline$c=2$, $K=2$ & $-\sqrt[3]{\frac{1}{2}}\simeq -0.7937$ &$0.534$ & $0.631$ \\
\hline
\end{tabular}
\caption{The paramagnetic and ferromagnetic
  solutions are stable to the left and right of $\rho_{P}^{S}$ and $\rho_{F}^{S}$ respectively. The stability of the paramagnetic solution is found analytically (see section \ref{sec:sta}). The point in which the ferromagnetic solutions first
  appears within the RS approximation $\rho_{F}$, as well as the stability of this solution, are found numerically by solving the RS self consistent equation (\ref{eq:messpas}) for different $\rho$.}\label{tab:values}
\end{table}

The picture that emerges from these calculations is the following: 
above $\rho_{F}^{S}$, the ferromagnetic RS solution is stable to 1RSB perturbations and thermodynamically relevant. Below this point, both  spin-glass solutions are unstable. On the other hand, the stability of the trivial paramagnetic solutions (P) depends on the parity of the loops. In Husimi lattices with even loop sizes ($c+1=4$, $c+1=6$, etc) this phase is never stable, while in lattices with odd loops sizes, this phase becomes stable below $\rho_{P}^{S}$.

\subsection{1RSB Solutions}

For each Husimi lattice and for each value of $\rho$, the 1RSB free energy $\phi(\mu)$ is
computed using the usual method of population dynamics. The 1RSB
prediction for the energy of the model is given by
$U_{1RSB}=\phi(\mu^*)$ where $\mu^*$ maximizes $\phi$. On the other
hand, the limit $\mu\rightarrow 0$ must recover the RS equations and 
$\phi(0)$ must coincide with one of the replica symmetric solutions.

In figures \ref{fig:otherHusimiUvsrho2} and \ref{fig:otherHusimiUvsrho}, $U_{1RSB}=\phi(\mu^*)$ and $\phi(0)$ are
shown simultaneously for the square and pentagon Husimi lattices with $K=1$. In both types of graphs, the ferromagnetic solution is thermodynamically relevant for large $\rho$, as our intuition tells us. In the case of odd loop sizes, the situation is similar to the one found in the triangular Husimi lattice: the trivial paramagnetic solution seems to be the thermodynamically relevant for negative enough $\rho$.

\begin{figure}[!htb]
\begin{center}              
            \includegraphics[scale=0.8,angle=0]{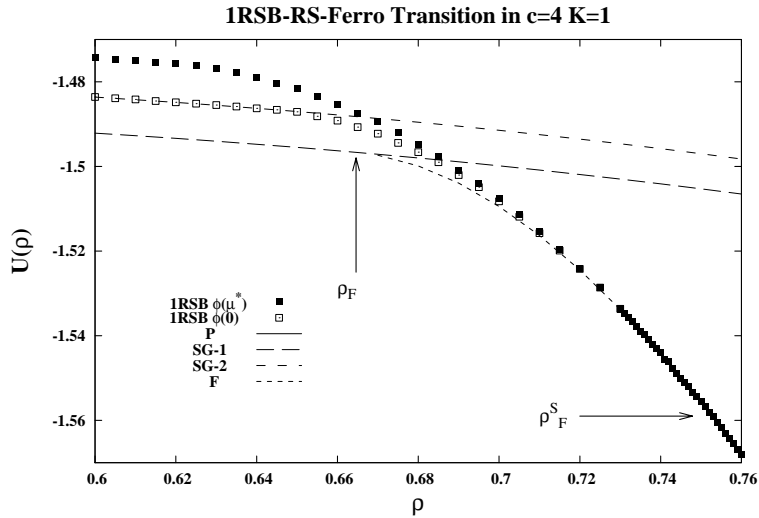} 
           \end{center}
	\caption{Energy versus $\rho$ close to the \textit{1RSB-Ferromagnetic} transition. The real thermodynamic behavior of the system is defined by the black squares. The replica symmetric ferromagnetic phase is not only stable above $\rho_F^S$, but also thermodynamically relevant, as the 1RSB solution smoothly becomes RS (see eq. \ref{eq:RSfrom1RSB}) as $\rho\rightarrow \rho^S_F$.}
	\label{fig:transferro}
\end{figure}

Summarizing, below the ferromagnetic stable region $\rho_{F}^{S}<\rho\leq 1$, the model has always a thermodynamically relevant 1RSB solution. In Husimi lattices with even loop sizes ($c+1=4$, $c+1=6$, etc.) this phase goes till $\rho=-1$, while in Husimi graphs with odd loop sizes and depending on K, this 1RSB phase may disappear for negative enough $\rho$ and in this case, the thermodynamics of the system is controlled by a trivial paramagnetic solution. However, it is worth mentioning that for large $cK$, $\rho_P^S \rightarrow -1$. Moreover, we will show below that the 1RSB solution may extend below $\rho_P^S$, so, in more general cases, we expect that this solution becomes thermodynamically relevant also for $\rho=-1$ \cite{luca}.

\begin{figure}[!htb]
\begin{center}              
            \includegraphics[scale=0.8,angle=0]{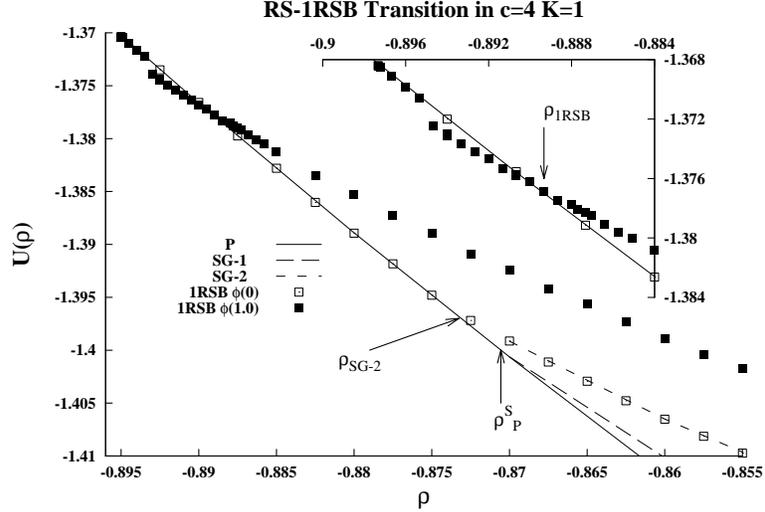} 
           \end{center}
	\caption{Energy versus $\rho$ close to the  \textit{1RSB-Paramagnetic} transition. The real thermodynamic behavior of the system is defined by the black squares, except for the unstable region in which the adiabatic continuation of the 1RSB solution goes below the paramagnetic curve (see caption). For $\rho<\rho_{1RSB}$ the thermodynamically relevant order parameter is the RS trivial paramagnetic solution. The transition \textit{1RSB-Paramagnetic} is not continuous: the structure of the order parameter changes abruptly when crossing $\rho_{1RSB}$. }
	\label{fig:transpara}
\end{figure}

To understand the characteristics of the transitions between the RS to the 1RSB phases we take the $c=4, K=1$ case as a model. We made a close-up and a careful calculation around the \textit{1RSB-Ferromagnetic} and the \textit{1RSB-Paramagnetic} transitions. They appear in figures \ref{fig:transferro} and  \ref{fig:transpara}, respectively. In both cases the simulations were done adiabatically, starting from the 1RSB zone and slowly varying $\rho$ towards the transition points and working always at $\mu^*$. In this way we guarantee that the algorithm finds (if it exits) the 1RSB solution.

As the figure \ref{fig:transferro} suggests, the \textit{1RSB-Ferromagnetic} transition is continuous. Increasing $\rho$,  a close inspection of the structure of the order parameter $\mathcal{Q}[Q(u)]$ (represented by the population of distributions $Q(u)$) shows that the amount of deltaic distributions $Q(u)=\delta_{u,u_0}$ grows smoothly as $\rho$ approaches $\rho_{F}^{S}$.

On the other hand (see figure \ref{fig:transpara}), the  \textit{1RSB-Paramagnetic} transition is quite different. There is region  $\rho_{1RSB} < \rho < \rho_P^S$ in which  the  thermodynamically relevant solution is 1RSB, but where the RS solution is stable under 1RSB perturbations.  Then, below the point $\rho_{1RSB} \sim -0.89$ the paramagnetic solution 
has higher energy and becomes thermodynamically relevant. Note on the inset that below this point there is still a region where the algorithm finds a non-trivial 1RSB solution of lower energy. This suggests that in this zone 1RSB solutions may appear but are exponentially rare. Then, by further decreasing $\rho$, the population dynamics finds only the trivial paramagnetic solution. The inspection of the distribution shows that this transition is discontinuous. Above $\rho_{1RSB}$  the order parameter $\mathcal{Q}_{1RSB}[Q(u)]$ is free of deltaic distributions $Q(u)=\delta_{u,u_0}$ and below $\rho_{1RSB}$ the solution is completely paramagnetic.

\subsection{The limit of the Bethe lattice}

Finally, we focused on the unbiased model $\rho=0$. In this case, all the Husimi lattices, with the exception
of the triangular ($c=2$ and $K=1$), are at least 1RSB (see eq. (\ref{eq:rhoparam})). 

If we refer to the Husimi
lattice considering only the local connectivity of the sites (number
of neighbors), then, a Husimi lattice with degree $K+1$ 
is locally very similar to a Bethe lattice with degree $k+1=2(K+1)$.
In table \ref{tab:GSE} we present a few ground state energies at the 1RSB level of approximation for different Husimi lattices and compare them with similar results for Bethe lattices. Note that for $c=7, K=1$ and $c=5, K=2$, the 
differences between the Bethe and the Husimi ground states are already very small.

\begin{table}[!htb]
\begin{tabular}{l @{\hspace{1cm}} c @{\hspace{5mm}} r @{\hspace{5mm}}}
\begin{tabular}{|l @{\hspace{1cm}}|c @{\hspace{5mm}} |}
\hline Husimi      &$U_{GS}$ \\
\hline$c=2$, $K=1$ & $-\frac{4}{3}=-1.333$ \\
\hline$c=3$, $K=1$ & $-1.444\pm 0.002 $\\
\hline$c=5$, $K=1$ & $-1.468\pm 0.002 $\\
\hline$c=7$, $K=1$ & $-1.470\pm 0.002 $\\
\hline Bethe, $k=3$, $c=\infty$,$K=1$ & $-1.471\pm 0.002 $\\
\hline
\end{tabular} & &\begin{tabular}{|l @{\hspace{1cm}}|c @{\hspace{5mm}} |}
\hline Husimi      &$U_{GS}$ \\
\hline$c=2$, $K=2$ & $-1.471\pm 0.002$ \\
\hline$c=3$, $K=2$ & $-1.819\pm 0.002$ \\
\hline$c=5$, $K=2$ & $-1.823\pm 0.002$ \\
\hline Bethe, $k=5$, $c=\infty$,$K=2$ & $-1.825\pm 0.002$ \\
\hline
\end{tabular}
\end{tabular} 
\caption{Ground state energies within the 1RSB approximation of
  different Husimi lattices  for $\rho=0$.} \label{tab:GSE} 
\end{table}

\begin{figure}[!htb]
\begin{center}              
            \includegraphics[scale=0.8,angle=0]{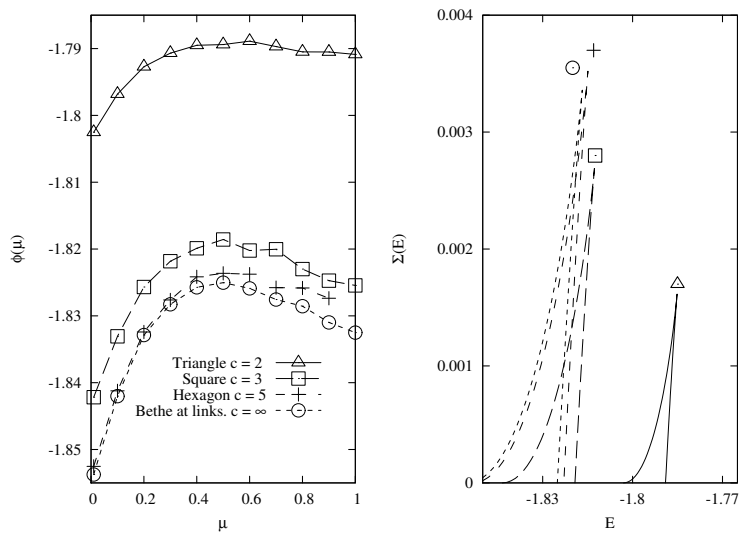} 
           \end{center}
	\caption{Free energy $\phi(\mu)$ (left) and complexity
          $\Sigma(\epsilon)$ (right) of
        different Husimi lattices at  $\rho=0$ and $K=2$. The graphics for $K=1$ are similar. Also
        represented the EA solution in the Bethe lattice with degree $k=5$. The symbols in the right
      panel correspond to the legend in the left panel. The early coincidence ($c=5$) between our short loop model in Husimi graphs, and the short-loop-free model in the Bethe lattices, gives a thermodynamic threshold for the relevance of loops.}
	\label{fig:phiK2}
\end{figure}

This similarity goes beyond the ground state calculations. In figure \ref{fig:phiK2} we plot
the  free energy $\phi(\mu)$, and the complexity
$\Sigma(\epsilon)$  for different Husimi lattices with
$K=2 $ and $\rho=0$ and for the Bethe lattice.  As expected, the figure shows a clear
maximum at $\phi(\mu^*)$ and the two usual branches in the $\Sigma(\epsilon)$
plot\cite{MP2}. From these plots, we may also conclude that the local similarity between 
Bethe and Husimi lattices, for which the relation $k+1=2(K+1)$ holds,  turns out to be true,  on thermodynamic grounds for
loop sizes greater than $c+1=6$ ($c+1=8$ if $K=1$). This suggest that,  at least for the EA model, the Bethe
approximation may work even under conditions less restrictive than
loop sizes of order $\ln{N}$.
 
\section{Conclusions}
\label{sec:dis}

We solved the Edwards-Anderson model in a Husimi graph at $T=0$. 
We presented closed analytical (RS) expressions
as a function of $\rho$ for
the existence and the stability of a trivial paramagnetic solution
(P). For the triangular Husimi lattice ($c=2$ and $K=1$) we obtained similar expressions
for the appearance and the stability of two spin-glass solutions (SG-1, SG-2) and a ferromagnetic solution
(F). For other cases, these points were calculated numerically (see
table \ref{tab:values}).
Within the 1RSB approximation we obtained, using a population dynamics algorithm,
the value of the ground state energies of different lattices as a
function of $\rho$ (see figures \ref{fig:otherHusimiUvsrho2} and 
\ref{fig:otherHusimiUvsrho}).

The main picture emerging from this work is that
for Husimi lattices with even loop sizes, the system is at least 1RSB
in all the range  $-1<\rho<\rho_F^S(c,K)$. Above $\rho_F^S(c,K)$ the
model has a RS ferromagnetic solution, and the transition from the 1RSB phase to the ferromagnetic one is continuous.
On the other hand, for lattices with 
 odd loop sizes  a trivial
paramagnetic solution is stable under 1RSB perturbations up to $\rho_{P}^{S}=-\left(\frac{c
K-2}{c K}\right)^{\frac{1}{c+1}}$, in addition for small $K$ this solution may become thermodynamically relevant below $\rho_{1RSB} \leq \rho_{P}^S$ where the order parameter jumps discontinuously from a non-trivial 1RSB solution towards trivial paramagnetic distributions. 
For the particular case of the triangular lattice with $K=1$ it was proven that the trivial paramagnetic solution is actually the unique and thermodynamically relevant solution (at least in the 1RSB frame) for all $\rho<0$. 

Finally, we focused on the $\rho=0$ case and computed the ground state energies for  different values of $c$
and $K$ (see table \ref{tab:GSE}). Our results suggest that the energy and complexity of the EA model in Husimi
lattices with loop sizes $c+1 \geq 8$ are already well described by an equivalent short-loop-free EA model in a Bethe lattice. This, in turn, can be considered a thermodynamic threshold for the \textit{shortness} of loops, less restrictive than the usual $\sim \log(N)$.

\section{Acknowledgments}

We thank F. Ricci-Tersenghi and M. Pretti for useful discussions and comments. We
also acknowledge the support of the NET-61 from the ICTP, and the ICTP for kind hospitallity.

\section{Appendix A}\label{appendix}

Let us prove that the trivial paramagnetic solution is actually the only thermodynamically relevant solution of the self-consistent equation (\ref{eq:selfconsQQ}) for the triangular Husimi lattice ($c=2$ and $K=1$), in the interval ($-1\leq\:\rho\:<\rho_P^S=0$). In doing so, we will follow exactly the same procedure used to study the instability of the Replica Symmetric solutions. We will presume that a non-trivial solution exists for the self-consistent equation, and we will show that it is unstable to small variations of the probability of paramagnetic distributions of messages $Q(u)=\delta_{u,0}$.

Let us call $\mathcal{Q}[Q]$ the solution of the self-consistent equation at 1RSB level of approximation. A perturbed order parameter would be $\mathcal{Q}[Q]+\pi \delta^F(Q(u)-\delta_{u,0})$. If we write down the self-consistent equation (\ref{eq:selfconsQQ}) using this order parameter, and keep to the first order in $\pi$ in the right-hand side, we will get the following equation for the evolution of the perturbation weight:
\begin{equation}
\pi'\delta_{u,0}= 2 \pi \mathrm{E}_J \int \delta^{(F)}(Q-\hat{Q}[J_{0,1},J_{1,2},J_{2,0},Q,\delta_{u,0}]) \ud\mathcal{Q}[Q] 
\end{equation}
where $\mathrm{E}_J$ is the expectation over the coupling constants $J$, and the 2 multiplying the integral is a combinatorial factor. The possible realizations of the disorder can be grouped into two sets: those with an odd number of antiferromagnetic interactions, like $(-1,-1,-1)$ and $(-1,1,1)$; and those with an even number like $(1,1,1)$ and $(-1,-1,1)$. Elements of the first group occur with a probability
\begin{equation}
P_{\mbox{frustration}}=(\frac{1+\rho}{2})^3+3(\frac{1+\rho}{2})^2(\frac{1-\rho}{2})=\frac{1-\rho^3}{2}
\end{equation}
while the other group has a complementary probability. Whenever the disorder happens to be frustrated (first group), the convolution  $\hat{Q}[\underline{J},Q,\delta_{u,0}]=\delta_{u,0}$ regardless of the actual distribution $Q$. On the other hand, if the disorder around the triangle is not frustrated (second group) $\hat{Q}[\underline{J},Q,\delta_{u,0}]=Q(u)$, thus reproducing the probability distribution of $\mathcal{Q}[Q(u)]$. Then the contribution to the perturbation is only given by the frustrated triangles, and we get for the iteration:
\begin{equation}
\pi'= 2 \pi P_{\mbox{frustraci\'on}}
\end{equation}
The instability condition would be given by:
\[2 P_{\mbox{frustration}}=1-\rho^3>1
\]
which happens to be true in the interval $-1\leq\rho<0$. This proves that any order parameter is instable to paramagnetic perturbations. This leaves no room for other stable order parameters than the trivial paramagnetic one, concluding our proof.

\bibliography{bibliografia}
\end{document}